\documentclass[journal]{IEEEtran}
\usepackage{amsmath}
\usepackage{graphicx}
\usepackage{subfigure}
\usepackage{graphicx}
\usepackage{epstopdf}
\usepackage{verbatim}
\usepackage{color}

\ifCLASSINFOpdf

\else

\fi

\begin{document}

\title{Accurate Closed-Form Approximations to Channel Distributions of RIS-Aided Wireless Systems}

\author{Liang Yang, Fanxu Meng, Qingqing Wu, Daniel Benevides da Costa, and Mohamed-Slim Alouini

\thanks{L. Yang and F. Meng are with the College of Computer Science and Electronic Engineering, Hunan University, Changsha
410082, China (e-mail:liangy@hnu.edu.cn, mengfx@hnu.edu.cn).}

\thanks{Q. Wu is with the State Key Laboratory of Internet of Things for Smart
City and Department of Electrical and Computer Engineering, University of Macau, Macao, 999078, China (e-mail:qingqingwu@um.edu.mo).}

\thanks{D. B. da Costa is with the Department of Computer Engineering,
Federal University of Cear¨¢, Sobral, CE, Brazil (email: danielbcosta@ieee.org).}

\thanks{M.-S. Alouini is with the CEMSE Division, King Abdullah University of
Science and Technology (KAUST), Thuwal 23955-6900, Saudi Arabia (e-mail:
slim.alouini@kaust.edu.sa).}}

\maketitle

\begin{abstract}
This paper proposes highly accurate closed-form approximations to channel distributions
of two different reconfigurable intelligent surface (RIS)-based wireless system setups,
namely, dual-hop RIS-aided (RIS-DH) scheme and RIS-aided transmit (RIS-T) scheme. Differently from previous
works, the proposed approximations reveal to be very tight for arbitrary number $N$ of reflecting
metasurface's elements. Our findings are then applied to the performance analysis of the
considered systems, in which the outage probability, bit error rate, and average channel
capacity are derived. Results show that the achievable diversity orders $G_d$ for RIS-DH
and RIS-T schemes are $N-1<G_d<N$ and $N$, respectively. Furthermore, it is revealed that
both schemes can not provide the multiplexing gain and only diversity gains are achieved.
For the RIS-DH scheme, the channels are similar to the keyhole multiple-input multiple-output
(MIMO) channels with only one degree of freedom, while the RIS-T scheme is like the transmit diversity structure.

\end{abstract}


\IEEEpeerreviewmaketitle

\section{Introduction}
Reconfigurable intelligent surfaces (RISs) are man-made surfaces
composed by electromagnetic (EM) materials, which is highly controllable
by leveraging electronic devices. Recent results indicate that RISs have great
potential due to their promising  gains achieved in terms of spectral
and energy efficiency without the need of higher hardware complexity and cost.

Owing to their promising gains, RISs have been extensively investigated
in the literature along the last year. In particular, in [1],
the authors proposed a practical phase shift model for the RISs. In [2], the authors
provided a comprehensive overview of the basic characteristics of the large intelligent
surface/antenna (LISA) technology, and listed various potential application examples.
In [3], the authors derived the well-known squared power gain of the RIS-assisted single-cell wireless system
and proved that it has better performance than conventional massive multiple-input
multiple-output (MIMO) systems as well as better performance than multi-antenna
amplify-and-forward (AF) relaying networks with smaller number of antennas, while
reducing the system complexity and cost. In [4], the authors studied the beamforming
optimization of RIS-assisted wireless communication under the constraints of
discrete phase shift. In [5], the authors used the RIS to improve the achievable
rate of unmanned aerial vehicle (UAV)-enabled communication system and proposed an optimization algorithm to
maximize the average achievable rate. In [6], the authors used an RIS in
dual-hop free-space optical and radio-frequency (FSO-RF) communication systems
to improve their performance. More recently, in [7], the authors studied the coverage and signal-to-noise
ratio (SNR) gain of RIS-assisted communication systems.

Although the previous works have provided interesting contribution to the field about RISs,
a common assumption in some of them [8] is that the non-central chi-square (NCCS) distribution
was used to approximate the channel distribution of the RIS-aided wireless schemes.
However, the NCCS distribution is based on the central limit theorem (CLT),
which means that it is only applicable to a large number of reflecting elements
$N$, i.e., $N\gg1$. Thus, for small values of $N$, it was shown that there is a big difference
between the analytical results and the simulation results (for instance, see Fig. 7 of [8]).
This limitation motivates our work. Specifically, relying on the small argument approximation
(SAA) for the sum of probability density functions (PDFs) used in [9], [10],
this paper proposes highly accurate closed-form approximations to the channel distributions of
RIS-aided systems assuming arbitrary values of $N$
\footnote{As far as the authors are aware, this
is the first time that general, accurate approximations are
proposed for a system setup assuming arbitrary number of
reflecting elements at the RIS, and our work can be used
as a benchmark for future studies in the field}.
More specifically, similar to [8], we consider two different RIS-based transmission scenarios,
namely, dual-hop communication (RIS-DH) case and the case when an
RIS is employed as a transmitter (RIS-T). For the former scenario, the RIS is
deployed between the source and the receiver and operates like a relaying system. For the second case,
the RIS is the own transmit source. More details about the two structures can be found in [8].
Based on the proposed approximations, closed-form expressions for the outage probability,
average bit error rate (BER), and average channel capacity are derived. Furthermore, an
asymptotic analysis is carried out. Results show that the achievable diversity orders $G_d$ for
RIS-DH and RIS-T schemes are $N-1<G_d<N$ and $N$, respectively. Furthermore, it is revealed that both schemes
can not provide the multiplexing gain and only diversity gains are achieved. For the RIS-DH scheme,
the channels are similar to the keyhole MIMO channels with only one degree of freedom [11]-[13], while
the RIS-T scheme acts like the transmit diversity structure.
\section{system and channel models}
In this section, we consider two different RIS-aided transmission
schemes and formulate their corresponding channel distributions.
\subsection{RIS-DH Case}
Consider an RIS-aided dual-hop communication which includes one source (S),
an RIS with $N$ reflecting elements, and one destination (D).
Similar to [8], it is assumed perfect knowledge of the channel state information 
(CSI). In addition, the RIS can provide adjustable phase shifts, which can be controlled 
and programmed by a software.

From [8], one can obtain the maximum SNR as
\begin{equation}
\begin{split}
\gamma_{1} = \frac{\left ( \sum_{i=1}^{N}  \alpha_{i}\beta _{i}\right )^2 E_{s1}}
{N_{0}} = R_{1}^2 \bar \gamma_{1},
\end{split}
\end{equation}
where $N_{0}$ is the noise power, $\alpha_{i}$ and $\beta _{i}$ are independent and identically
distributed (i.i.d.) Rayleigh random variables (RVs) with mean $\sqrt{\pi}/2$ and variance $(4-\pi)/4$, $E_{s1}$ is the average
power of the transmitted signals, and $\bar \gamma_{1} = E_{s1}/N_{0}$ denotes the average SNR.

In [8], the authors relied on the CLT to assume that $R_{1}$ is a Gaussian
variable so that $R_{1}^2$ follows the NCCS distribution.
However, under this assumption, the results of [8] revealed that the difference
between the analytical and simulation results is very large for small values of $N$.
To address this issue, next we make use of the idea presented in [9] to derive the
probability density function (PDF) of $R_{1}^{2}$.

Let $\chi_{i} = \alpha_{i}\beta _{i}$. Then, the PDF of $\chi_{i}$ can be
readily obtained as $f_{\chi_{i}}(r)=4rK_{0}(2r)$, where $K_{v}(\cdot)$ is the
modified $v$-order Bessel function of the second kind [14, Eq.(8.432)]. In addition, the PDF of a
random variable $\chi$ follows a $K_{G}$ distribution and can be written as
\begin{equation}
f_{\chi}\left ( r \mid m,k,\Omega \right )=\frac{4\Psi ^{k+m}}{\Gamma \left ( m \right )
\Gamma \left ( k \right )}r^{k+m-1}K_{k-m}\left ( 2\Psi r \right ),
\end{equation}
where $\Omega \overset{\triangle }{=}E[\chi^{2}]$ is the mean power, $\Gamma(\cdot)$ is the Gamma
function [14, Eq.(8.310.1)], $\Psi=\sqrt{km/\Omega}$, $k$ and $m$ are the shaping parameters of the
distribution.

Comparing the PDF of $\chi_{i}$ with (2), one can observe that the PDF of
$\chi_{i}$ is a special case of the $K_{G}$ distribution
when $k=m=1$. In [9], it was stated that the PDF of the sum of multiple $K_G$
random variables can be well-approximated by the PDF of $\sqrt{W}$ with $W=\sum_{i=0}^{N}\chi_{i}^{2}$,
in which the PDF of $W$ is approximated by a squared $K_{G}$ distribution.
Therefore, the PDF of $R_{1}$ can be formulated as
\begin{equation}
f_{{R}_{1}}\left ( r \right ) {=}
\frac{4\Xi^{k_{w}+m_{w}}r^{k_{w}{+}m_{w}{-}1}}{\Gamma \left ( k_{w} \right )\Gamma
\left ( m_{w} \right )}K_{k_{w}{-}m_{w}}\left ( 2\Xi r  \right ),
\end{equation}
where $k_{w}=\frac{-b+\sqrt{b^2-4ac}}{2a}$ and $m_{w}=\frac{-b-\sqrt
{b^2-4ac}}{2a}$ are the the shaping parameters, $\Xi = \sqrt{k_{w}m_{w}
/\Omega _{w}}$, and $\Omega _{w} = \mu _{R_{1}}(2)$
is the mean power of $R_{1}$. Moreover, the parameters $a$, $b$, and $c$ have been
defined in [9] and their values are related to the moment $\mu _{R_{1}}(n)$ of $R_1$, namely,
\begin{align}
\mu _{R_{1}}(n)&= \sum_{n_{1}=0}^{n}\sum_{n_{2}=0}^{n_{1}}...\sum_{n_{N-1}=0}^
{n_{N-2}}\binom{n}{n_{1}}\binom{n_{1}}{n_{2}}...\binom{n_{N-2}}{n_{N-1}} \nonumber\\
 &\times \mu _{\chi_{1}}(n-n_{1})\mu _{\chi_{2}}(n_{1}-n_{2})...\mu _{\chi_{N-1}}(n_{N-1}),
\end{align}
where $\mu _{\chi_{i}}(n)  = \Gamma^2(1+n/2)$ is the $n$th moment of $\chi_{i}$. Notice that $k_{w}$ and $m_{w}$ are real numbers.
For $b^2-4ac \leq 0$ corresponding to $k_{w}$ and $m_{w}$ being conjugate complex numbers,
$k_{w}$ and $m_{w}$ are set to the estimated modulus values of the conjugate complex number.

From (1) and (3), the PDF of $\gamma _{1}$ can be represented by a squared $K_{G}$ distribution.
Thus, the PDF of $\gamma_{1}$ can be written as
\begin{equation}
f_{\gamma _{1}}\left ( \gamma  \right )\approx \frac{2\widetilde{\Xi }^{k_{w}+m_{w}}\gamma ^{\left ( \frac{k_{w}+m_{w}}{2}-1
\right )}}{\Gamma
\left ( k_{w} \right )\Gamma \left ( m_{w} \right )}K_{k_{w}-m_{w}}\left ( 2\widetilde{\Xi }\sqrt{\gamma } \right ),
\end{equation}
where $\widetilde{\Xi }=\sqrt{k_{w}m_{w}/\left ( \overline{\gamma}_{1} \Omega _{w}  \right )}$.
Moreover, the cumulative distribution function (CDF) of $\gamma_{1}$ can be obtained as
\begin{equation}
F_{\gamma _{1}}\left ( \gamma  \right ) \approx \frac{1}{\Gamma \left ( k_{w} \right )\Gamma \left
( m_{w} \right )}G_{1,3}^{2,1}\left [ \widetilde{\Xi } ^{2}\gamma |_{k_{w},m_{w},0}^{1} \right ],
\end{equation}
where $G_{p,q}^{m,n}\left [ \cdot \right]$ is the Meijer $G$-function defined in [14, Eq.(9.301)].
To the best of the authors' knowledge, the above expressions are new in an RIS-aided wireless
systems context and can be useful in future studies due to their tightness, as shown later in numerical results section.

To obtain more insights, next we develop another approximate PDF expression by using the
fact $\left ( \sum_{i=1}^{N}  \alpha_{i}\beta _{i} \right )^2 \leq \left ( \sum_{i=1}^{N}
\alpha_{i}^2 \right ) \left ( \sum_{i=1}^{N}\beta_{i}^2 \right )$ [14, Eq.(11.112)] as
\begin{equation}
f_{\gamma _{1}}\left ( \gamma  \right )\approx \frac{2\gamma^{N-1}}
{\Gamma (N)^2\bar\gamma_{1}^N}K_{0}\left(2\sqrt{\frac{\gamma}{\bar\gamma_{1}}}\right).
\end{equation}

It is noteworthy that (7) equals to the PDF expression of the keyhole channels when we set
$n_T=n_R=N$ in [11]-[13].
Thus, in the RIS-DH scheme, the channels may be degenerated due
to the so-called keyhole effect [11]-[13], while the average SNR in [11]-[13] is
$\bar\gamma_{1}/N$ due to the MIMO structure.

Notice that (7) may not be tight enough, but
using it to evaluate the system performance can reveal the channel characteristics of the RIS-DH scheme.
\subsection{RIS-T Case}
Consider now the system model which includes an
RF signal generator, an RIS, and one destination (D), where
the RIS is used as a transmitter along with the RF signal
generator
Similar to [8], the RIS is deployed very close to the RF
signal generator, which implies that the channel attenuation between them is negligible.
Therefore, we use the RIS and RF signal generator together as a transmitter (T).
More details regarding this model can be found in [8].

From [8], the maximum instantaneous SNR at D can be expressed as
\begin{equation}
\begin{split}
\gamma_{2} = \frac{\left ( \sum_{i=1}^{N} \varepsilon _{i}\right )^2 E_{s2}}{N_{0}}
= \frac{R_{2}^2 E_{s2}}{N_{0}},
\end{split}
\end{equation}
where $E_{s2}$ is the power of the unmodulated signal, $\varepsilon _{i}$ is
a Rayleigh RV with mean $\sqrt{\pi}/2$ and variance $(4-\pi)/4$, and $R_{2}$ is
the sum of $N$ i.i.d. Rayleigh RVs $\varepsilon_{i}$, $i{=}1,...,N$.
Similarly, according to the CLT,
$R_{2}^2$ is assumed to follow a NCCS distribution [8].
However, such an assumption has only limited applications in practice. Fortunately, from [10],
the PDF of $R_{2}^2$ can be approximated by
\begin{equation}
\begin{split}
f_{R_{2}^2}(r) \approx \frac{r^{N-1}e^{-\frac{r}{B}}}{B^N(N-1)!},
\end{split}
\end{equation}
where $B = 1+(N-1)\Gamma^2(\frac{3}{2})$.

Thus, the PDF of $\gamma_{2}$ can be written as
\begin{equation}
\begin{split}
f_{\gamma_{2}}(\gamma) \approx \frac{\gamma^{N-1}e^{-\frac{\gamma}{B\bar \gamma_{2}}}}{(B\bar \gamma_{2})^N(N-1)!},
\end{split}
\end{equation}
where $\bar \gamma _{2} = E_{s2}/N_{0}$ is the average SNR.
Moreover, the CDF of $\gamma_{2}$ can be obtained as
\begin{equation}
F_{\gamma _{2}}(\gamma ) \approx 1-e^{-\frac{\gamma}{B\bar\gamma_{2}}}\sum_{k=0}^{N-1}\frac{\gamma^k}{(B\bar\gamma_{2})^k k!}.
\end{equation}

\section{Performance analysis}
In this section, based on the proposed approximations, we derive new
closed-form expressions for the outage probability, average BER,
and average channel capacity of the considered RIS-aided wireless systems.
\subsection{Outage Probability Analysis}
The reliability of the communication link is usually assessed by using the outage probability, which is defined as
\begin{equation}
\begin{split}
P_{out}=\rm Pr(  \gamma \leq  \gamma_{th}),
\end{split}
\end{equation}
where $\rm Pr(\cdot)$ indicates probability.
\subsubsection{RIS-DH Case}
From (6), we have
\begin{align}
P_{out1}=F_{\gamma_{1}}(\gamma_{th}).
\end{align}
Above result does not indicate the diversity order information. Therefore, it is necessary to
develop an asymptotic analysis for the RIS-DH scheme. As mentioned above, (7) is similar to
the PDF of the keyhole MIMO channels. Thus, from [11],[12], the diversity order of the RIS-DH
system is readily given by $N-1< G_{d}< N$, which is  equal to the multi-antenna systems over keyhole channels.
\subsubsection{RIS-T Case}
Similarly, from (11), the outage probability of the RIS-T case can be obtained as
\begin{align}
P_{out2}=F_{\gamma_{2}}(\gamma_{th}).
\end{align}
For high SNRs, utilizing $1-e^{-x}\sum_{s=0}^{N}\frac{x^s}{s!} \leq \frac{x^N}{N!}$ [15], we have
\begin{equation}
\begin{split}
P_{out2} \approx \frac{\gamma_{th}^N}{(B\bar\gamma_{2})^N N!},
\end{split}
\end{equation}
which indicates that the achievable diversity order $G_d$ is $N$.
\subsection{Average BER}
BER is a typical metric for evaluating the accuracy of data transmission.
For different binary modulation schemes, the unified BER expression is given by [16]
\begin{equation}
P_{e} = \frac{q^p}{2\Gamma (p)}\int_{0}^{\infty }\exp(-q\gamma )\gamma ^{p-1}F_{\gamma}(\gamma )d\gamma,
\end{equation}
where $F_{\gamma}(\gamma )$ is the CDF of $\gamma$, and the
parameters $p$ and $q$ denote different modulation
schemes, such as $p$ = 1 and $q$ = 1 for differential phase
shift keying (DPSK).
\subsubsection{RIS-DH Case}
From (6) and (16), and using [17, Eq.(07.34.21.0088.01)], we have
\begin{equation}
\begin{split}
P_{e1} = \frac{1}{2\Gamma (p)\Gamma (m_{w})\Gamma (k_{w})}G_{2,3}^{2,2}\left
( \frac{\Xi^2 }{\bar\gamma_{1}q} \vert \begin{matrix}
1-p,1\\
k_{w},m_{w},0
\end{matrix} \right ).
\end{split}
\end{equation}
Similar to the outage analysis, we also can obtain the diversity order of the RIS-DH system as $N-1< G_{d}< N$.
\subsubsection{RIS-T Case}
Similarly, from [14, Eq.(3.326.2)] and making use of (11) and (16), the BER for the RIS-T case can be derived as
\begin{equation}
\begin{split}
P_{e2} = \frac{1}{2}-\frac{q^p}{2\Gamma (p)}\sum_{k=0}^{N-1}\frac{1}{(B
\bar\gamma_{2})^k k!}\frac{\Gamma (p+k)}{(q+\frac{1}{B\bar\gamma_{2}})^{p+k}}.
\end{split}
\end{equation}

For high SNRs, similar to (15),we have
\begin{equation}
\begin{split}
P_{e2} \approx \frac{\Gamma (p+N)}{2\Gamma (p)(Bq\bar\gamma_{2})^{N}N!}.
\end{split}
\end{equation}

One can observe from (19) that the achievable diversity order $G_d$ is also $N$.
\subsection{Average Channel Capacity}
From [18], the average channel capacity can be evaluated in terms of the CDF as
\begin{equation}
C = \frac{1}{\ln(2)}\int_{0}^{\infty }(1+\gamma)^{-1}F_{c}^{\gamma}(\gamma)d_{\gamma},
\end{equation}
where $F_{c}^{\gamma}(\gamma) = 1 - F_{\gamma}(\gamma)$ is the complimentary CDF function.
\subsubsection{RIS-DH Case}
By using [17, Eq.(07.34.21.0086.01)] together with (6) and (20), the average channel capacity can be obtained as
\begin{align}
C_{1}{=}\frac{1}{\Gamma (m_{w})\Gamma (k_{w})\ln(2)}G_{2,4}^{4,1}
\left ( \frac{\Xi^2 }{\bar\gamma_{1}} \vert \begin{matrix}
0,1\\
0,0,k_{w},m_{w}
\end{matrix} \right ).
\end{align}
Similar to [13, Eq.(23)], applying the Jensen's inequality, we obtain a tight upper bound for $C_1$ as
\begin{align}
C_{1} &\leq \log_{2}(1+\Omega _{w} \bar\gamma_{1}).
\end{align}
Similar to asymptotic analysis in [13, Eq.(38)], $C_{1}$ can be asymptotically expressed as
\begin{equation}
C_{1} \approx \log_{2}\left ( \bar\gamma_{1} \right )+2 \log_{2}(e)\psi (N),
\end{equation}
where $\psi (N)= \Gamma'(N)/\Gamma (N)$ is the Euler's digamma function.
Like the observations in [13], one can see that the RIS-DH scheme does not provide the spatial
multiplexing gain and only diversity gain is achieved.
\subsubsection{RIS-T Case}
Similarly, by using [14, Eq.(3.353.5)] together with (11) and (20), the average channel capacity
of the RIS-T case can be expressed as
\begin{align}
C_{2} =& \sum_{k=0}^{N-1}\sum_{m=1}^{k}\frac{ (m-1)!(-1)^{k-m}
(B\bar\gamma_{2})^{m}}{(B\bar\gamma_{2})^k k!\ln(2)} \nonumber\\
&+\sum_{k=0}^{N-1}\frac{(-1)^{k-1}e^{\frac{1}
{B\bar\gamma_{2}}}E_{i}\left ( \frac{-1}{B\bar\gamma_{2}}\right )}{(B\bar\gamma_{2})^k k!\ln(2)},
\end{align}

where $E_{i}(\cdot)$ is the exponential integral function [14, Eq.(8.211)]. Similarly, we can
obtain the upper bound on $C_{2}$ as
\begin{align}
C_{2} &\leq \log_{2}(1+BN\bar\gamma_{2}).
\end{align}

Furthermore, we have
\begin{align}
C_{2} &\approx \log_{2}( \bar\gamma_{2})+ \log_{2}(BN).
\end{align}

Above formula indicates that the RIS-T scheme also only provides the diversity gain.

\section{Numerical and Simulation Results}
In this section, we provide numerical and simulation results to verify our analysis.
The parameter is set to $\gamma_{th}=20$ dB.

For the BER performance, we consider the DPSK modulation.
\begin{figure}[t]
\centering
\includegraphics[width=8cm,height=6cm]{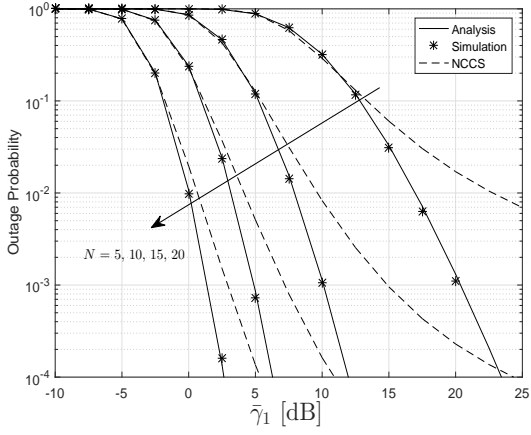}
\caption{Outage probability versus $\bar{\gamma}_{1}$ for the RIS-DH scheme.}
\end{figure}
\begin{figure}[h]
\centering
\includegraphics[width=8cm,height=6cm]{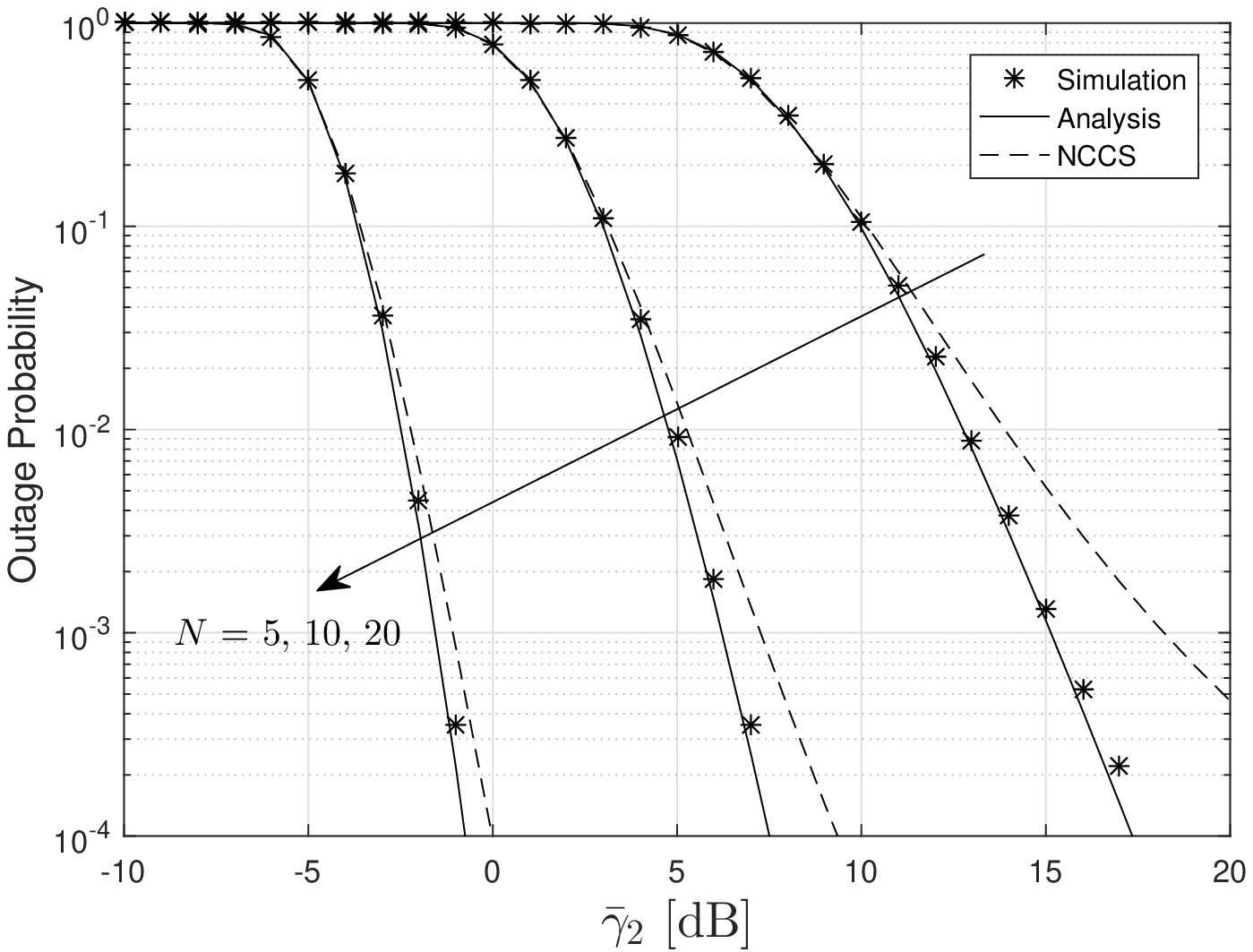}
\caption{Outage probability versus $\bar{\gamma}_{2}$ for the RIS-T scheme.}
\end{figure}
\begin{figure}[h]
\centering
\includegraphics[width=8cm,height=9cm]{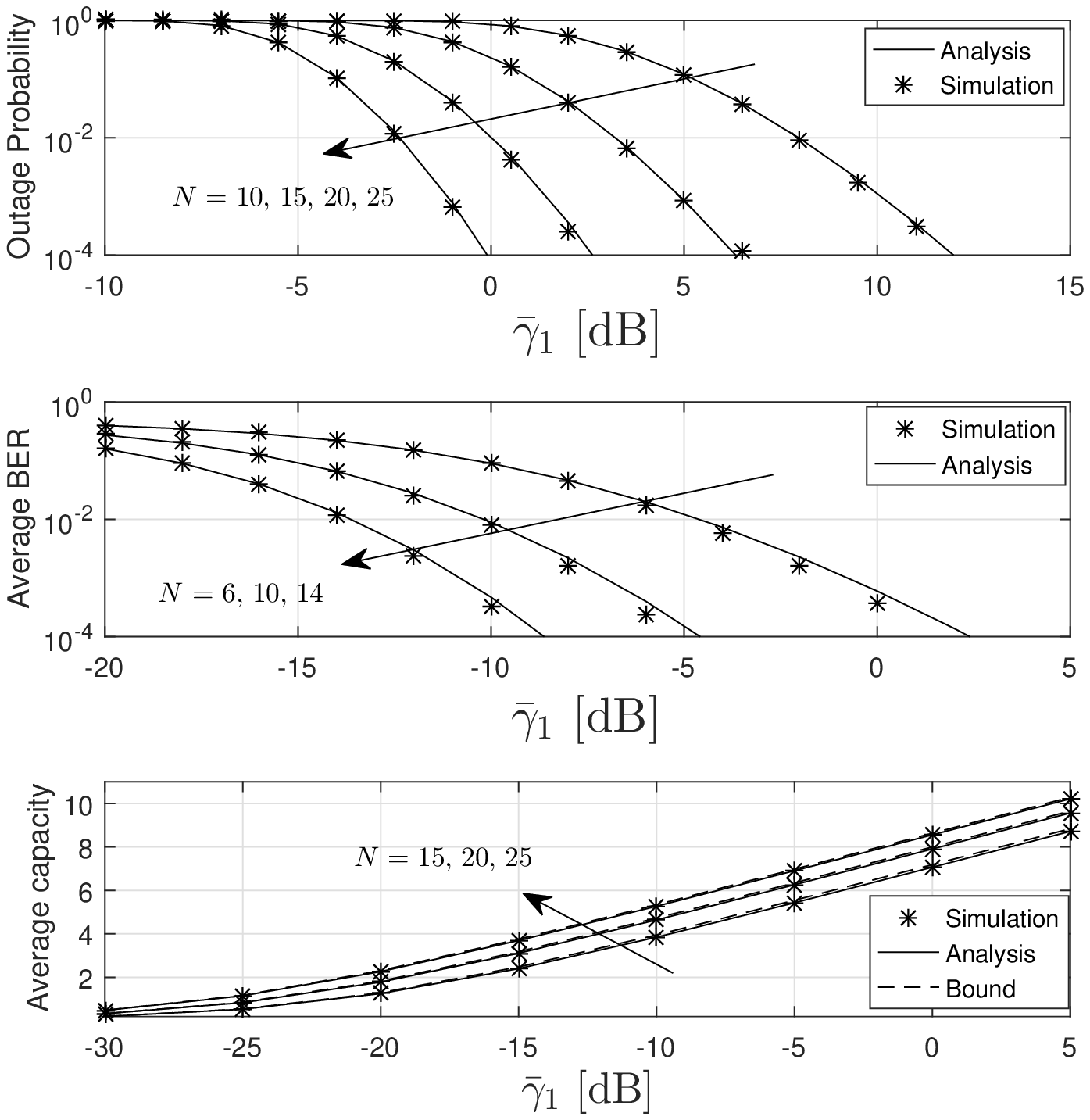}
\caption{Outage probability, average BER, and average capacity for the RIS-DH scheme.}
\end{figure}

\begin{figure}[h]
\centering
\includegraphics[width=8cm,height=9cm]{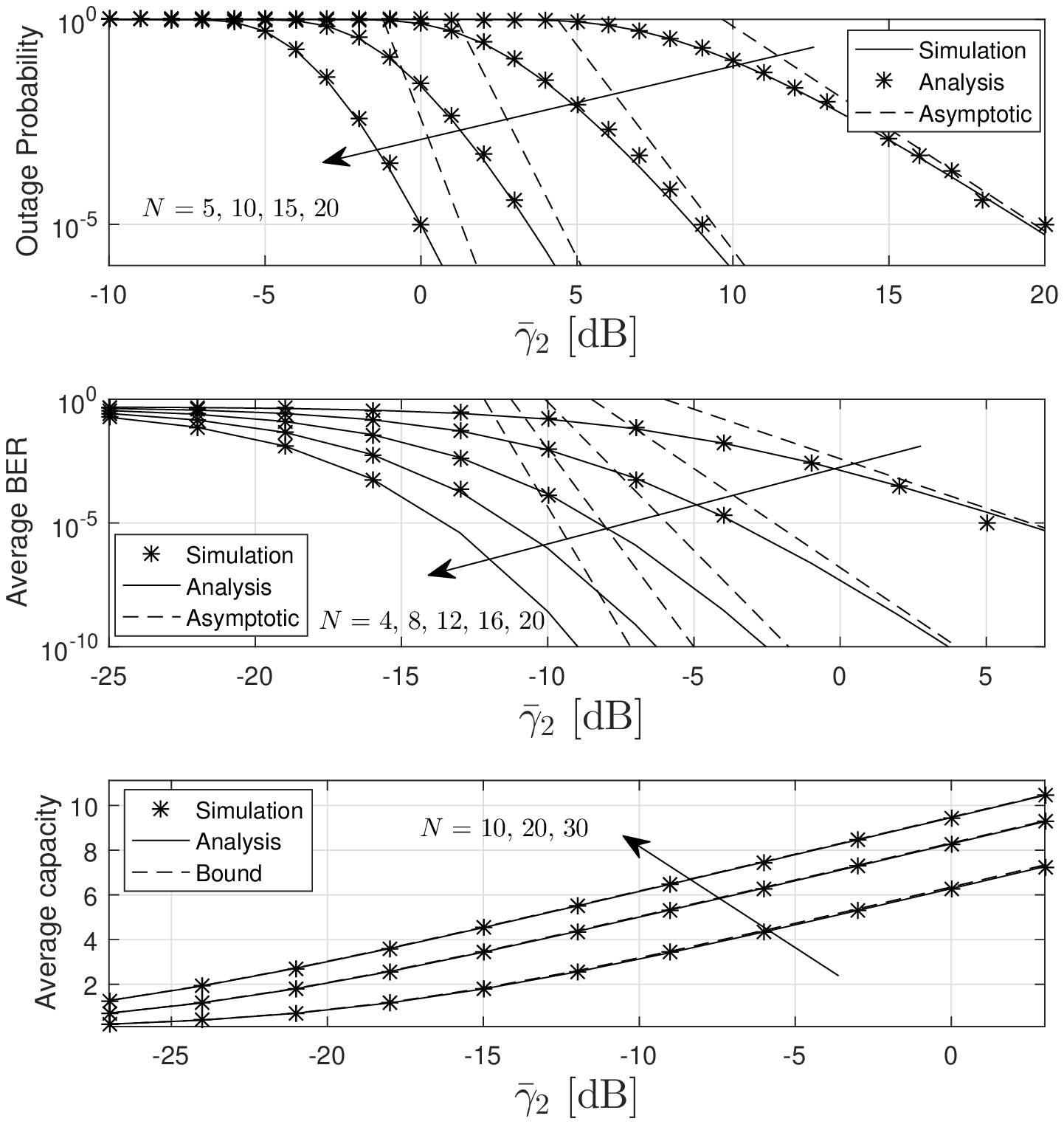}
\caption{Outage probability, average BER, and average capacity for the RIS-T scheme.}
\end{figure}
\begin{figure}[h]
\centering
\includegraphics[width=8cm,height=6cm]{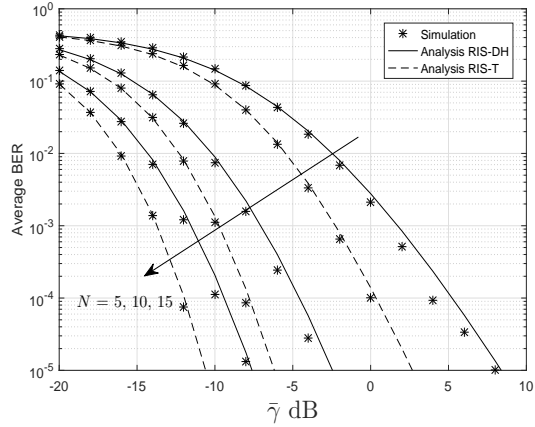}
\caption{BER curves $\bar{\gamma}$ for both schemes.}
\end{figure}
In Fig. 1, assuming the RIS-DH scheme, we compare the outage probability between the NCCS distribution and the squared $G_K$ distribution.
Notice that the outage probability using the NCCS distribution can be
evaluated by $P_{out3} = 1- Q_{\frac{1}{2}}\left ( \frac{\sqrt{\lambda}}{\sigma},
\sqrt \frac{{\gamma_{th}}}{\bar \gamma_{1} \sigma^2} \right )$, where $\lambda = \frac{N^2 \pi^2}{16}$ and $\sigma^2 = N(16-\pi^2)/16$.
When $N$ is small, using NCCS results
in a large error. Thus, for the RIS-DH case, it is obvious that using the squared
$G_K$ distribution is more accurate compared to the NCCS distribution.

In Fig. 2, assuming the RIS-T scheme, we compare the outage probability between the
NCCS distribution and the approximation provided in (9). Notice that the outage
probability using the NCCS distribution can be
evaluated through $P_{out3}$ by setting $\lambda = \frac{N^2 \pi}{4}$ and
$\sigma^2 = N(4-\pi)/4$.
Similar to Fig. 1, our proposed approximate approach can get more accurate
results than the NCCS distribution. Thus, from both Figs. 1 and 2, one can
clearly see that the proposed approximate distributions for both transmission
scenarios are very precise for the whole average SNR range, while the NCCS distribution is too
loose. Note that for high values of $N$, our approach still works very well.

In Fig. 3, we plot the curves for the outage probability, average BER, and average channel
capacity for the RIS-DH case. It is clearly observed that the analytical
results are highly consistent with the simulation results. Furthermore, one can see
that the upper bound of the average channel capacity is very tight.

In Fig. 4, the curves for outage probability, average BER, and average capacity
are depicted for the RIS-T case. As expected, the simulation results are consistent with the analytical
results. Similar observations obtained in Fig. 3 apply here as well. From both Figs. 3 and 4,
one can see that the slope of the capacity curves are kept to a constant regardless of
the value of $N$. The reason is that both schemes only provide diversity gain.

In Fig. 5, by setting $\bar\gamma_{1}=\bar\gamma_{2}=\bar\gamma$, we plot the BER curves
for both schemes. As expected, it is observed that the RIS-DH scheme has a smaller diversity
order compared to the RIS-T scheme, which verifies our asymptotic analysis.

\section{Conclusion}
In this work, we developed two novel statistical distributions to approximate the channel
distributions of the RIS-aided wireless systems which generalize existing result based on
NCCS distribution. Then, outage probability, average BER,
and average capacity were derived in closed-form. Simulation results showed that the proposed distributions
perform better than the NCCS distribution. Moreover, it was shown that the achievable diversity
orders $G_d$ for RIS-DH and RIS-T schemes are $N-1<G_d<N$ and $N$, respectively. Furthermore,
both of the schemes can not provide the multiplexing gain and only diversity gains are achieved.
As a potential future work, line-of-sight scenarios that are commonly modeled by Rician fading
can be taken into account, in which a totally new approximate approach would need to be proposed.

\end{document}